\documentclass[twocolumn,preprintnumbers,amsmath,amssymb,aps,prl]{revtex4}
\usepackage{setspace}
\usepackage{graphicx} 
\usepackage{amsmath}
\usepackage{color}

\begin{document}

\title{Directed collective motion of bacteria under channel confinement}

\author{Hugo Wioland$^1$\footnote{Present address: Institut Jacques Monod, Centre Nationale pour la Recherche Scientifique (CNRS), UMR 7592, Universit\'e Paris Diderot, Sorbonne Paris Cit\'e, F-75205 Paris, France}, Enkeleida Lushi$^2$ and Raymond E. Goldstein$^1$}
\affiliation{$^1$ Department of Applied Mathematics and Theoretical Physics, Centre for Mathematical Sciences, University of Cambridge, Wilberforce Road, Cambridge CB3 0WA, UK\\
$^2$ School of Engineering, Brown University, 182 Hope Street, Providence, Rhode Island 02912, USA}

\date{March 3, 2016}

\begin{abstract}
Dense suspensions of swimming bacteria are known to exhibit collective behaviour arising from the interplay of 
steric and hydrodynamic interactions.
Unconfined suspensions exhibit transient, recurring vortices and jets, whereas those confined
in circular domains may exhibit order in the form of a spiral vortex. 
Here we show that confinement into a long and narrow macroscopic 
`racetrack' geometry stabilises bacterial motion to form a 
steady unidirectional circulation.
This motion is reproduced in simulations of discrete swimmers that reveal the crucial role that
bacteria-driven fluid flows play in the dynamics.
In particular, cells close to the channel wall produce strong flows which advect cells in the bulk 
against their swimming direction.
We examine in detail the transition from a disordered state to persistent directed motion as a function of the channel width,
and show that the width at the crossover point is comparable to the typical correlation length of swirls seen in the
unbounded system.
Our results shed light on the mechanisms driving the collective behaviour of bacteria and other active matter systems, 
and stress the importance of the ubiquitous boundaries found in natural habitats.
\end{abstract}

%\noindent{\it Keywords\/}: locomotion, Bacillus subtilis, active matter, micro-swimmer suspension, biofluid dynamics, microbiological flows.

\maketitle

\section{Introduction}
Spreading and survival of populations of bacteria often depend on their ability to behave collectively: 
cells aggregate into swarms to seek and migrate towards nutrient-rich regions~\cite{BenJacob2004, Ariel2013}, 
organise into biofilms resistant to antibiotics~\cite{Mah2001, Wilking2013}, respond to 
starvation by building fruiting bodies~\cite{Shimkets1999, Branda2001} or opt for cannibalism~\cite{Gonzales2011}.
In such organisations, the surrounding environment often plays a major role, 
through its chemical composition~\cite{Mah2001, Ariel2013} or geometrical constraints~\cite{Wilking2013, Grant2014}.
A complex and fascinating issue is how the various chemical or mechanical interactions between the microorganisms and 
their environment can guide the intricate dynamics of populations. 

Theoretical approaches to this question may utilize methods from the emerging field of
`active matter', which has taken motivation from collective behaviour in suspensions of flagellated bacteria
(e.g., \textit{Escherichia coli} and \textit{Bacillus subtilis})
~\cite{Dombrowski2004, Cisneros2011, Sokolov2012, Aranson2013, Gachelin2013, Wioland2013, Lushi2014, Wioland2016} as well 
as those composed of molecular motors and the filaments along which they move~\cite{Nedelec,Schaller}.
Using discrete~\cite{discrete}, continuum~\cite{SaintillanShelley}, and phenomenological models, 
this growing field of research has studied self-organisation of 
populations of interacting motile organisms or other kinds of active and driven objects, 
often giving rise to striking collective behaviours.
Recent studies have predicted that physical confinement can
have a strong impact on the spatio-temporal organization, and indeed may result in unidirectional flows 
~\cite{Voiturez2005, Fielding2011, Woodhouse2012, Ravnik2013, Neef2014, Tsang2016}.
Experimental realisations of confined active matter however remain relatively rare~\cite{Wioland2013, Lushi2014, Wioland2016, Bricard2013, Kumar2014, Vladescu2014, Bricard2015}.

The interactions between swimming cells in bacterial suspensions have two main components: direct steric repulsion and 
long-range hydrodynamic forces created by the action of multiple flagella. 
These interactions lead to complex collective motion: while direct repulsion between rod-like bacteria 
produces local alignment akin to nematic liquid crystal ordering, hydrodynamic forces 
can advect and reorient nearby bacteria~\cite{Ramaswamy,PedleyKessler,Drescher2011} and
power macroscopic turbulent 
patterns such as jets and swirls much faster than an individual bacterium~\cite{Dombrowski2004,Cisneros2011, Sokolov2012}.
Yet, these collective structures are not permanent, instead they are {\it recurrent} and {\it transient}.

Previous experiments have been successful in controlling the bacterial migration through gradients of 
chemoattractants~\cite{BenJacob2004, Ariel2013} or by varying external flow and environment properties~\cite{Rusconi2014}.
An alternative that we consider here takes advantage of interfaces, as can be commonly encountered
in bacterial habitats like soil.
Recent studies have explained how a single microorganism interacts with surfaces~\cite{Drescher2011, Kantsler2013}, 
yet little has been done to experimentally study the collective dynamics of motile bacteria and 
other types of active matter under confinement~\cite{Wioland2013, Lushi2014, Wioland2016, Bricard2013, Bricard2015}.
Different interfaces (fluid, solid) or confinement topologies (e.g. planar, linear, circular) 
could lead to various macroscopic organisations.

\begin{figure}
  \centering
  \includegraphics[]{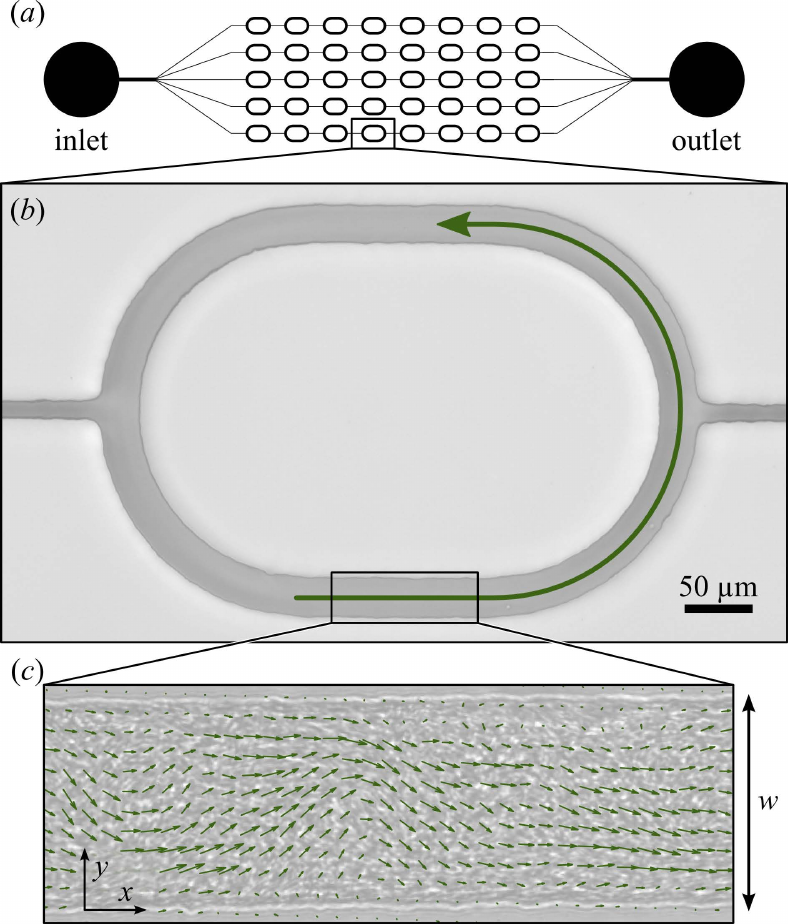}
  \caption{\label{setup} Experimental setup. (a) Layout of the entire microfluidic circuit.  Bacteria are injected through 
  the inlet, which is sealed prior to recording. (b) Detail of one racetrack in the PDMS chamber 
device. Arrow indicates one of two equivalent directions for spontaneous circulation. (c) close-up of
  the circulation. Arrows: instantaneous bacterial flow field as measured by PIV.}
\end{figure}

Here we describe an experimental setup able to stabilise a dense suspension of \textit{Bacillus subtilis} into a persistent stream
through physical confinement alone.  As shown in figure \ref{setup},
bacteria are introduced into an array of a thin periodic millimetre-long racetracks, allowing for quantification of many realizations 
of the resultant flow patterns.  Our chief finding is that there exists a clear transition from
a `turbulent' state to stable circulation
for channel widths $\lesssim 70\,\mu$m, or a scale comparable to correlation length of cell orientations, or equivalently the diameter of the characteristic swirls seen in effectively unbounded suspensions~\cite{Dombrowski2004,Cisneros2011,Dunkel2013}.
Numerical studies of a discrete model of swimmers which incorporates both steric and hydrodynamic interactions between the cells reveal that the suspension motion is dominated by the bacteria-driven fluid stream. 
In particular, cells close to the bounding walls create a strong fluid flow
which advects the cells in the bulk of the channel against their swimming direction in a manner similar to that seen in circular drops of suspension \cite{Wioland2013,Lushi2014}.
Analysis of the suspension kinetics sheds light on the mechanism that drives bacterial collective behaviour.
This organisation shows strong connections with previous work on bacterial pumping in microfluidic 
chambers~\cite{Kim2008} and cell swimming or alignment against a shear fluid flow
~\cite{Rusconi2014, Hill2007, Kaya2012, Marcos2012, Kantsler2014}.

%%%%%%%%%%%%%%%%
\section{Materials and Methods}
\subsection{Experiments}
Dense suspension of motile \textit{B. subtilis} (wild-type, strain 168) were prepared from 
an overnight culture, diluted $200$-fold into fresh Terrific Broth (TB, Sigma) and incubated at 
$35^\circ$C for $\approx5$h. When the bacteria reached a high motility fraction ( $> 90\%$ motile cells), $10\,$ml of the 
suspension was concentrated by centrifugation ($1500\,$g, $10\,$min), and the 
pellet was used in the experiments without further dilution. We estimate the cell volume fraction to be $\sim20\%$.\\

Polydimethylsiloxane (PDMS) microchambers were fabricated with racetracks 
($20\,\mu$m high, $20-130\,\mu$m wide and more than a $1$ mm long, figure \ref{setup}(a)), linked 
by thin inlets ($\approx10\,\mu$m wide). Bacteria were introduced into the chamber with a syringe and the main inlets 
were closed prior to recording to avoid any unwanted motion (visual inspection verified the absence of
any net flow between racetracks).

Each preparation was allowed to reach a steady state over two minutes and videos lasting $5$ seconds were acquired
with a high-speed camera (Fastcam, Photron, $125$ frames/sec) under bright field illumination, 
using a $63\times$ oil-immersion objective (Zeiss).  The velocity field of the bacteria was determined by Particle 
Image Velocimetry (using a customised version of mPIV~\cite{mpiv}) with no time averaging (figure \ref{setup}(b)). 
This method yields the properties of the bacterial motion, not the associated fluid flow.

\subsection{Measuring the bacterial orientation}
%PIV itself yields only the local velocity field of the bacteria, not their head-tail orientation.
Bright field images yield only the local velocity field of the bacteria and their rough alignment~\cite{Wioland2013}, not their head-tail orientation.
To determine the relationship between swimming and motion directions 
we labelled the flagella and membrane of a subset of the population with two different fluorophores,
as described previously~\cite{Lushi2014}. The mutant 
strain amyE::hag(T209C) DS1919 3610~\cite{Guttenplan2013} was grown in the same conditions as wild-type bacteria and labelled 
following the protocol of Guttenplan {\it et al.}~\cite{Guttenplan2013}. $1\,$ml 
of the suspension was centrifuged ($1000\,$g, $2\,$min) and resuspended in $50\,\mu$L of 
Phosphate Buffered Saline (PBS) containing $5\,\mu$g/mL Alexa Fluor 488 $\textrm{C}_5$ Maleimide 
(Molecular Probes) and incubated at room temperature for $5\,$minutes. This fluorophore reacts 
with the cysteine added to the flagella protein flagellin (gene \textit{hag}). Bacteria were then 
washed in $1\,$mL PBS and resuspended in PBS containing $5\,\mu$g/mL FM 4-64 (Molecular 
Probes). This fluorophore incorporates itself into lipid bilayers and labels cell membranes. The 
excess of fluorophore was removed in a final wash and the bacteria were resuspended into 
$50\,\mu$l PBS. A fraction of the labelled mutants was then gently mixed with unlabelled wild-type 
bacteria. Cells cannot swim in PBS, which lacks a carbon source, 
but recover their initial motility once transferred into Terrific Broth.

Images were acquired on a spinning disc confocal microscope (Zeiss Axio Observer Z1, camera 
Photometrics Evolve 512 Delta) at $6$ fps. Both 
fluorophores were excited with a $488\,nm$ laser and the emission was filtered with a GFP filter
(barrier filter $500-550\,nm$, Zeiss) for Alexa Fluor 488 $\textrm{C}_5$ Maleimide and DsRed filter 
(barrier filter $570-640\,nm$, Zeiss) for FM 4-64. We took a sequence of three images: membrane 
(false coloured red), flagella (false coloured green) and again membrane (false coloured blue, 
figure \ref{fluo}) and measured the bacterial motion from the membrane displacement 
between the first and last image. We then deduced the orientation of the cell (its swimming 
direction) from the bundling of the flagella (at the rear of the cell) 
and from the average position of the membrane relative to this bundle.

\subsection{Modelling and Simulations}
We used numerical simulations, adapted from a recent method~\cite{Lushi2014}, to understand
the self-organisation of the suspension.
Bacteria are represented as motile ellipses, subject to steric and hydrodynamic interactions and
confined into a periodic channel. 
Each swimmer generates a dipolar ``pusher''-flow field and is affected by the fluid flow disturbances  
created by the other swimmers.

We use a 2D domain with periodic boundaries of length $L$ and 
width $2w$ where $y=0,w$ are the channel walls. 
Each swimmer is modelled as an ellipse of length $\ell=1$, width $\ell/4$, described
by its centre of mass $\mathbf{X}_i$ and orientation $\mathbf{P}_i$. 
$\mathbf{X}_i$ and $\mathbf{P}_i$ are randomly initiated and follow the dynamics~\cite{Lushi2014,Lushi2013}:
\begin{align}
\partial_t \mathbf{X}_i &= U_0\mathbf{P}_i + \mathbf{v} +  \Xi^{-1}_i\sum_{j \neq i} \mathbf{F}^e_{ij}\label{xdot}~, \\
\partial_t \mathbf{P}_i &= (\mathbf{I} - \mathbf{P}_i\mathbf{P}_i^T)(\gamma \mathbf{E} + \mathbf{W}) \mathbf{P}_i + k\sum_{j \neq i}
(\mathbf{T}^e_{ij} \times \mathbf{P}_i).\label{pdot}
\end{align}
Equation (\ref{xdot}) describes self-propulsion with constant speed (chosen as $U_0=1$) 
along the cell direction $\mathbf{P}_i$, advection by the fluid velocity $\mathbf{v}$ interpolated 
at the swimmer position, and pairwise steric repulsion with force $\mathbf{F}^e_{ij}$ between 
swimmers. Here, $\Xi = m_{||}\mathbf{P}_i\mathbf{P}_i^T + m_{\perp}(\mathbf{I} - \mathbf{P}_i\mathbf{P}_i^T)$ 
with the mobility parameters $m_{\perp}$=$2 m_{||} \approx 2$. The first term of equation (\ref{pdot}) 
is Jeffery's equation and describes rotation of the particle by the fluid flow $\mathbf{v}$ with 
$2\mathbf{E}=\nabla \mathbf{v}+\nabla \mathbf{v}^T$,  $2\mathbf{W}=\nabla \mathbf{v}-\nabla \mathbf{v}^T$; 
$\gamma\approx 0.9$ for ellipses with aspect ratio 4. 
The last term of equation (\ref{pdot}) describes swimmer rotation due to torques from steric interactions with 
neighbours with $k\approx5$. The purely repulsive steric forces $\mathbf{F}^e_{ij}$ 
and torques $\mathbf{T}^e_{ij}$ between swimmers are obtained using the
method described by Constanzo {\it et al.}\cite{Constanzo2012}. Each swimmer is discretized 
into $n_b=4$ beads that interact with other swimmers through a soft capped Lennard-Jones 
potential~\cite{Lushi2014}. This potential allows some overlaps but does not over-restrict the time-stepping. Noise 
terms are not included in the dynamics. 

\begin{figure}[htps]
	\centering
  	\includegraphics[]{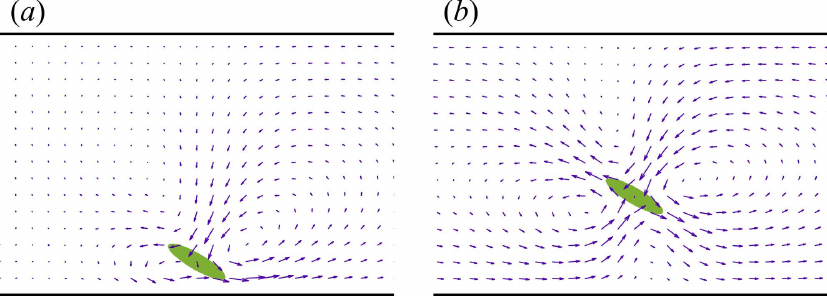}\\
  	\caption{\label{oneswimmer} Examples of the fluid flow generated by one swimmer inside a channel (a) at the boundary and (b) in the bulk.}
\end{figure}

To approximate the boundary condition at $y=0, w$, 
we use a system of images. For each swimmer $i$ in the channel 
$(x,y) \in ([0,L],[0,w])$, a mirror swimmer is placed in $(x,y) \in ([0,L],[0,-w])$ with centre of 
mass $\mathbf{X}_i-2(\mathbf{X}_i \cdot \mathbf{e}_y)\mathbf{e}_y$ and orientation 
$\mathbf{P}_i - 2(\mathbf{P}_i \cdot \mathbf{e}_y)\mathbf{e}_y$. We calculate the active stress 
tensor $\mathbf{S}^a_i$=$\sigma \mathbf{P}_i\mathbf{P}_i^T$ 
that results from locomotion for 
both the swimmers and their images. The non-dimensional stresslet strength is set to $\sigma = -1$ for a 
slender pusher swimmer with length $\ell=1$ and speed $U_0=1$~\cite{Lushi2013}. 
The fluid velocity $\mathbf{v}$ is obtained by solving with Fourier series 
the (non-dimensional) 2D Stokes Equations with the extra active stresses
\begin{eqnarray}\label{stokes}
-\nabla^2 \mathbf{v} + \nabla q = \nabla \cdot \sum_{i} \mathbf{S}^a_i \delta (\mathbf{x}-\mathbf{X}_i), \quad
\nabla \cdot \mathbf{v} = 0
\end{eqnarray}
in the entire (real and mirror) domain $(x,y) \in ([0,L],[-w, w])$ which is assumed bi-periodic. 
Using the mirror domain in the $y$-direction for solving equation \ref{stokes} is a 
computationally efficient way to approximate the effects of the PDMS interface 
as a no-stress boundary condition (figure \ref{oneswimmer}). Note that to guarantee a no-slip boundary condition for the fluid flow, other methods, e.g.~\cite{discrete} are more appropriate, but require higher computational power for many swimmer simulations.

 The free parameters in the simulations are the swimmer length $\ell:=1$, its shape parameter, its swimming speed $U_0:=1$, the stresslet strength $\sigma:=-1$ which determines the magnitude of the generated fluid flows, as well as the channel dimensions. We note that we did not match these with the experimental values, hence the comparisons between the results are qualitative and not quantitative.

\begin{figure*}
  \centering
  \includegraphics[width = 0.65\textwidth]{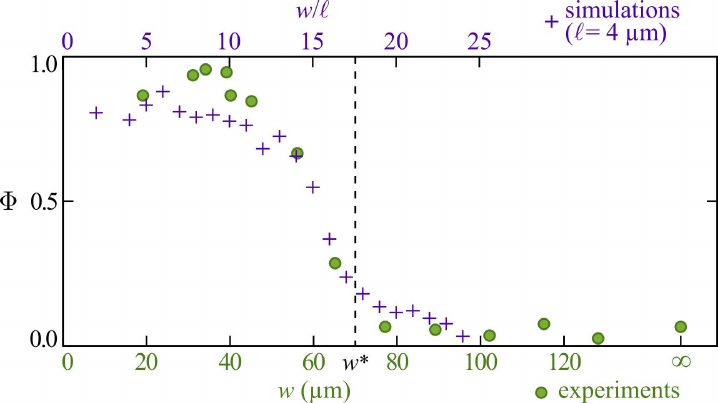}
  \caption{\label{circulation} Normalised net flow $\Phi$ 
  for a variety of channel widths in both experiments (circles) and simulations (crosses).
  Experimental points are averaged over at least five movies, recorded on 
  at least two different days, simulations are averaged over a time $5U_0/\ell$. 
  $\infty$: an unconfined quasi-2D chamber.}
\end{figure*}

%%%%%%%%%
\section{Results}
\subsection{Confinement stabilises a bacterial stream}
We inject a dense suspension of swimming \textit{B. subtilis} into $20\,\mu$m high racetracks.
When confinement is strong enough, the turbulent collective motion is stabilised and forms 
a persistent stream (figure \ref{setup}). 
This circulation is stable for tens of minutes until bacteria stop swimming due to oxygen depletion.  

\newpage
We quantify the overall motion by measuring the normalised net flow:
\begin{equation}
	\Phi = \left| \frac{ \sum \mathbf{u}\cdot\mathbf{e}_x }{\sum ||\mathbf{u}||} \right|,
\end{equation}
where $\mathbf{u}$ is the bacterial flow as measured by PIV, $\mathbf{e}_x$ the unit 
vector along the channel main direction (figure \ref{setup}(b)) and summing over all PIV 
sub-windows inside the channel and over $\approx70$ frames of a $5$ seconds long 
movie. We observed CW and CCW circulation with equal probability and therefore study 
the absolute value of $\Phi$.
$\Phi=1$ indicates a purely longitudinal flow (e.g. $\mathbf{u} = \|u\|\mathbf{e}_x$) 
and $\Phi=0$ indicates a globally stationary suspension.

Figure \ref{circulation} shows a clear dependence of the net flow $\Phi$ on the channel width $w$: 
while the suspension exhibits a strong circulation ($\Phi>0.7$) for thin chambers, it quickly 
transitions into a stationary turbulent state ($\Phi<0.2$) around $w^*_{\textrm{exp}}=70\,\mu$m.

We compared these results with simulations in a 2D periodic channel domain,
to mimic the straight section of the racetracks (figure \ref{setup}(b)).
Ellipsoid swimmers, subject to both steric and hydrodynamic interactions, self-organise 
into a stream similar to the observed experimental motion, as seen in figure \ref{simulations}(d).
We computed the net flow $\Phi$, this time summing over bulk swimmers only.
We found the same qualitative behaviour with strong circulation only for thin channels $w<w^*_{\textrm{sim}}=17\ell$ 
(where $\ell$ is the swimmer length).

Figure \ref{circulation} shows that the transition from a stable stream to an overall stationary state
can be matched between experiments and simulation ($w^*_{\textrm{exp}}=w^*_{\textrm{sim}}$) by choosing $\ell=4\,\mu$m. 
 Note that other outcomes, e.g. flow strength, are not necessarily matched.

\begin{figure*}
  \centering
  \includegraphics[width=6in]{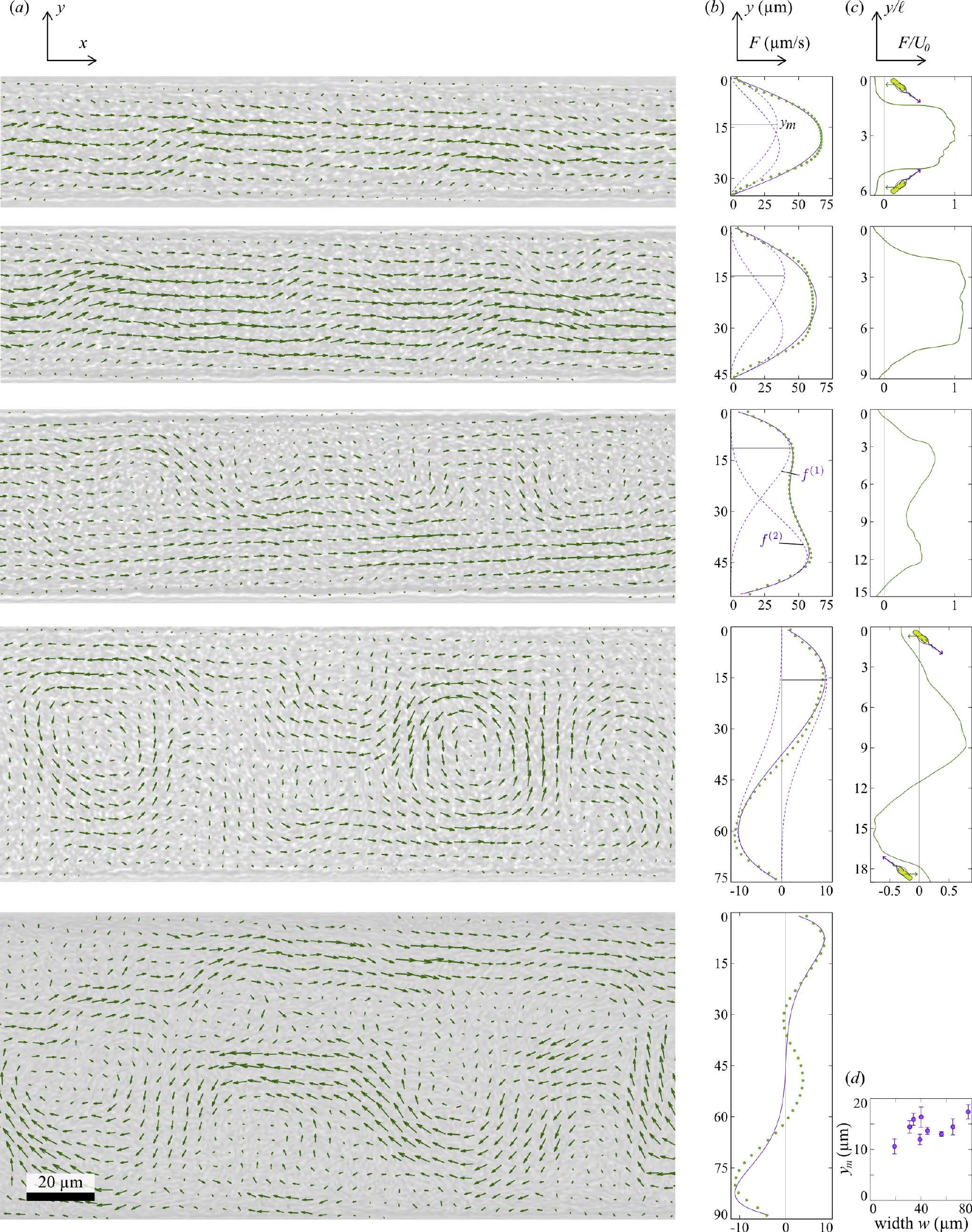}
  \caption{ \label{profile} 
  Bacterial flow pattern and profiles depending on the channels width.
  (a) Bacterial circulation measured by PIV for channels width $35,45,55,75,90\,\mu$m. 
  (b) Averaged bacterial flow profile $F(y)$ in experiments. 
  Measurements (dots) are fitted as the sum 
  (solid line) of two beta distribution functions,  $f^{(1)}_{\mathrm{beta}}$ and $f^{(2)}_{\mathrm{beta}}$ (dotted lines). 
  (c) Averaged bacterial flow profile $F(y)$ in simulations. 
  The distance is expressed in terms of swimmer length 
  $\ell$ and the fluid flow in terms of the single cell swimming speed $U_0$. Sketches of the swimmers
  at the edges indicate their orientations (green arrow) and the direction of the generated fluid flow (purple arrow). 
  (d) Distance $y_m$ between the PDMS interface and the maximum of the first beta distribution 
  function, as indicated in (b) by a thin black line. Each point is an average taken over at least 
  five movies, recorded on at least two different days.
  Error bars: standard error.}
\end{figure*}

%%%%%
\subsection{Bacterial flow patterns under confinement}
In unconfined environments, dense bacterial suspensions exhibit typical patterns of jets and swirls~\cite{Cisneros2011, Sokolov2012, Aranson2013, Gachelin2013}. 
PDMS interfaces in racetracks change this pattern in order to allow for the net circulation. 
We now study the pattern stabilisation, through the averaged flow profile, depending on the channel width.
If the flow were driven by a pressure gradient, a passive Newtonian fluid would produce a parabolic or 
pump flow~\cite{Brody1996}. Yet bacteria collectively generate irregular flow patterns.

We measure the profile or cross section of the flow patterns, averaged over time and space:
\begin{equation}
	F(y) = \left< \mathbf{u}(x,y,t)\cdot\mathbf{e}_x \right>_{(x,t)},
\end{equation}
where the direction of $\mathbf{e}_x$ is chosen such that $F(y)$ is on average positive. 
Figure \ref{profile}(a) shows different behaviours depending on the channel width $w$. 

For the thinnest channels ($w\approx35\,\mu$m) the flow is directed parallel to the channel length and $F$ takes a 
quasi-parabolic profile. $F$ then flattens for $w\approx45\,\mu$m and splits into two 
peaks for $w\approx55\,\mu$m. At such width, small swirls start to emerge but are not strong enough
to alter the circulation. 
When the width reaches the critical value $w>w^*_{\textrm{exp}}=70\,\mu$m, the suspension is able to form
full vortices. $F$ takes a sinusoidal form such that bacteria on each side move in opposite directions 
and most vortices circulate in the same direction (CCW in figure \ref{profile} (a)). 
For wider racetrack ($w\approx100\,\mu$m) the suspension recovers its quasi-turbulent state and 
$F$ exhibits several oscillations. In this last case, if we averaged over a longer period of time, 
the resulting profile should come to zero.

For channels between $50$ and $70\,\mu$m in width, we observe that the profile $F$ displays two maximum values, 
at a distance $y_m$ from the PDMS interface (figure \ref{profile}(b)). In order to measure this length
for different widths, we fit $F$ with a combination of two symmetric beta distribution functions:
\begin{align}
F_{fit}(y) &= f^{(1)}_{beta} + f^{(2)}_{beta} \nonumber \\
                    &=A_1(\frac{y}{w})(1-\frac{y}{w})^{\beta-1} +  A_2(1-\frac{y}{w})(\frac{y}{w})^{\beta-1},
\end{align}
where $A_1$ and $A_2$ are the amplitude of each function and $\beta$ is  
a parameters setting the shape of the distribution. Each beta function is the symmetric of 
the other except for the amplitude: $f^{(1)}_{beta}(y) = \frac{A_1}{A_2} f^{(2)}_{beta}(w-y)$. 
The beta distribution functions are merely used for simplicity and 
have no physical connection to the bacterial suspensions.

We found that $F_{fit}$ correctly describes the flow profiles until $w\approx75\,\mu$m 
(figure \ref{profile}(a)). 
For $w\approx35\,\mu$m, the two beta functions are almost identically and mostly overlap. They then come apart to form a 
double-peaked function ($w\approx55\,\mu$m) which then becomes anti-correlated ($A_1A_2<0$) 
for $w\approx75\,\mu$m. For the largest racetrack, $F_{fit}$ only captures the first and last peaks but 
not the bulk behaviour. 

We then measure the distance $y_m$ at which $f^{(1)}_{beta}$ reaches 
its maximum (represented as a thin line on figure \ref{profile}(b)). 
Figure \ref{profile}(d) shows that $y_m$ does not 
depend on the racetrack width: the flow profile is simply the sum of two beta functions, 
of width-independent shape, and that move apart as the channels widen,
suggesting that the circulation is driven by bacteria close to the wall.

Simulations show a similar behaviour, with parabolic ($w = 6\ell$), flattened ($w = 9\ell$), 
double-peaked ($w = 15\ell$) and sinusoidal profiles ($w = 19\ell$, figure \ref{profile}(b)).
The only significant difference lies in the bacteria motion at the boundary:
simulations reveal that these swimmers move against the bulk circulation: 
$F_{\mathrm{simu}}(0)\cdot F_{\mathrm{simu}}(3\ell)<0$.
We also observed experimentally that bacteria at interfaces swim against the bulk circulation, 
but PIV measurements could not catch this feature.

Both the swimmer motion at the boundary and the flow profile fitting suggest that    
bacteria close to the wall play an important role in the suspension ordering.

%%%%%%%%%%
\subsection{Boundaries and fluid flow drive the bacterial circulation}
\begin{figure*}
  \includegraphics[width=0.95\textwidth]{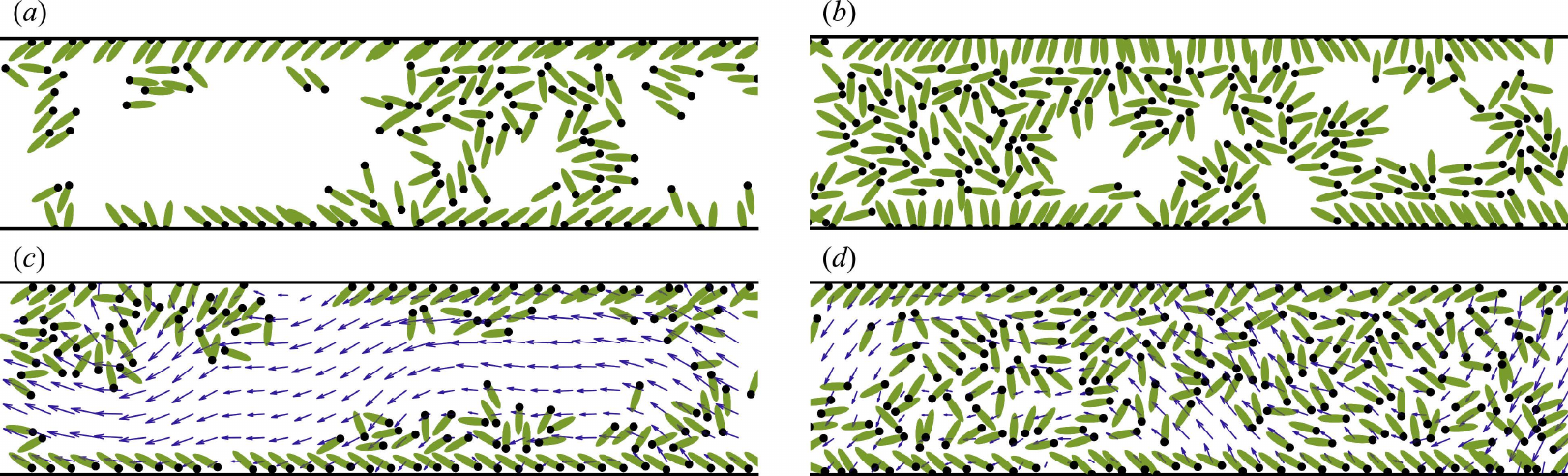}
  \caption{\label{simulations} Snapshots from simulations of micro-swimmers in a 2D channel. 
  (a, b) Dilute and dense suspensions of motile ellipses without hydrodynamical interactions ($\sigma=0$). 
  (c, d) Dilute and dense suspensions with hydrodynamical interactions ($\sigma=-1$). Collectively-generated 
  fluid flow fields are shown superimposed (the magnitude in plot (d) has been halved). 
  In both cases the fluid flow is not purely longitudinal and a wavy-like pattern is discernible.}
\end{figure*}

We turn to simulations to understand how boundaries affect the bacteria organisation and in particular to find 
the causes of the bulk circulation as observed in experiments.
We consider different scenarios: dilute or dense suspensions, with or without hydrodynamic 
interactions. The examples in figure \ref{simulations} have a channel width of $w = 6\ell$.

We first neglect hydrodynamics and consider swimmers interacting between themselves and with the boundary 
through direct steric repulsions only. 
In both dilute and dense cases (figure \ref{simulations}(a),(b)), swimmers aggregate at the boundary
to form packs or hedgehog-like clusters, as previously observed~\cite{Constanzo2012,Wensink2008}. 
Due to their elongated shape, swimmers at the boundary exhibit a local nematic alignment, 
with heads facing the interface, and slowly move alongside it.
Cells in the bulk jam in clusters with little organisation discernible.

\begin{figure*}[htps]
  \includegraphics[width=0.95\textwidth]{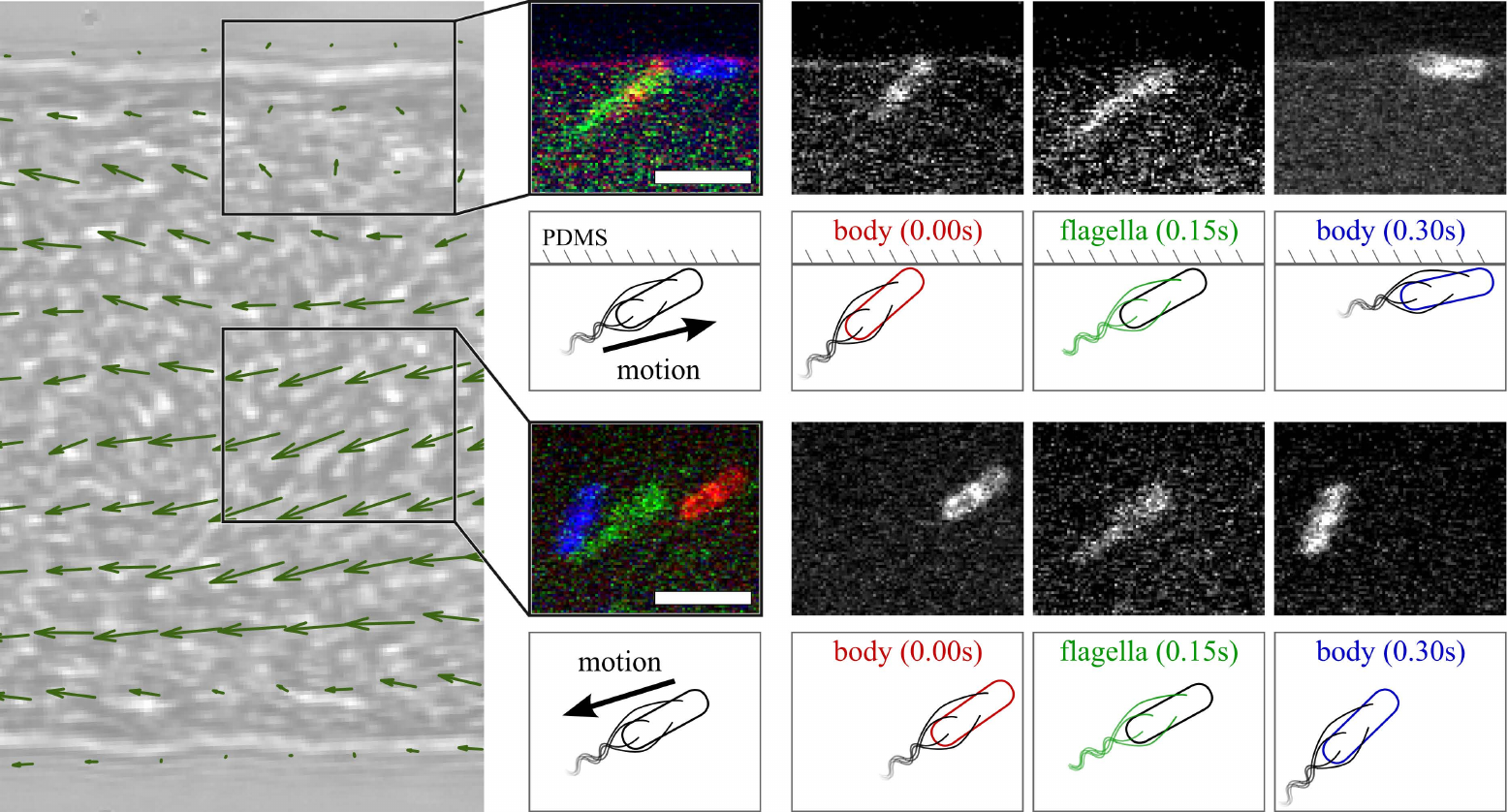}\\
  \caption{\label{fluo} 
  Fluorescently labelled bacteria indicate orientation and motion 
  direction of cells at the PDMS wall and in bulk. Mutant 
  \textit{B. subtilis} (strain amyE::hag(T204C) DS1919 3610~\cite{Guttenplan2013}) were 
  tagged at the body and flagella, respectively with FM 4-64 
  (false coloured red and blue) and Alexa 488 $\textrm{C}_5$ maleimide (false coloured green)}
\end{figure*}

We next include fluid flows generated by bacteria.
In dilute suspensions, as shown in figure \ref{simulations}(c), swimmers aggregate 
at the boundary and form packed layers similar to the steric-only case.
However they generate a strong fluid stream, against their swimming direction. 
If the channel width is sufficiently small, the fluid reorients the opposite layer 
such that swimmers on each channel side move in the same direction.
This packing and organisation is not observed experimentally at low densities, most probably because 2D simulations 
do not allow for overlap whereas bacteria in $20\,\mu$m deep racetracks can easily swim and roll-over over one-another. 

When the density is further increased (figure \ref{simulations}(d)), the same mechanism is at work:
the two boundary-bound layers of swimmers drive a strong fluid flow, reorienting and advecting the cells that
now fill up the centre of the channel. Bulk swimmers have a biased orientation against this flow 
but do not swim fast enough to overcome the advection. They are thus transported against the 
boundary layers giving rise to a macroscopic bulk circulation similar to the experimental observations 
(figure \ref{setup}(b)).

The simulations also clarify the profile shapes (figure \ref{profile}).
Sketches of the swimmers, drawn at the channel walls in figure \ref{profile}(c),
indicate their swimming orientation and the direction of the fluid flow they push back.
This flow advects over a distance $\sim y_m$ the cells in the bulk, whose overall collective motion 
can be fitted with the double beta function profiles.

Experimentally, bacteria can flow over $60\,\mu$m/s, 
significantly faster than the single cell swimming speed of $10\,\mu$m/s. 
This high velocity suggests that the bulk bacterial motion is indeed dominated by fluid advection.

To verify the organisation predicted by simulations, we track fluorescently labelled bacteria (figure~\ref{fluo}).
We deduced the cell orientation from the relative position of the body and flagella and the overall motion from the 
body displacement. We found the same biased orientation against the flow in the bulk of thin channels 
($w<w^*=70\,\mu$m): out of the $24$ labelled bacteria tracked,
$4$ were oriented along the $\mathbf{e}_y$ direction, $4$ were swimming with the flow and 
$16$ were swimming against the flow (oriented against their overall motion direction). 
Cells at the surface had a forward motion, against the bulk circulation, confirming the organisation found in simulations.

%%%%%%%%%%%%%%%%
\subsection{Spatial and temporal variations in bacterial flow patterns}
Our current model - edge bacteria driving the bulk flow through hydrodynamic interactions - explains
the overall circulation. As described earlier, bacteria in thin racetracks also form partial and travelling swirls, 
reminiscent of the unconfined turbulent state. 
We study this flow pattern by computing both the correlation lengths and the motion of swirls.

\begin{figure*}
  \centering
  \includegraphics[width=0.58\textwidth]{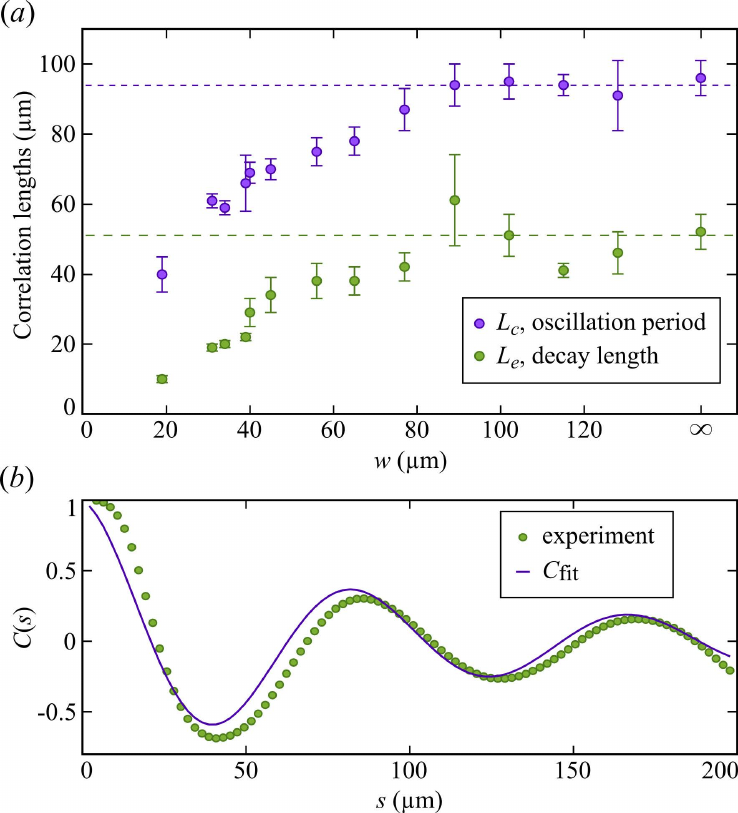}
  \caption{\label{graph} Bacterial flow correlation depending on the channel width $w$.
  (a) measurement of the correlation lengths depending on the racetrack width $w$. 
  Both oscillation period and amplitude decay length scale reach a plateau state at large width
  (dashed lines).
  Points are averaged over at least five movies, recorded on at least two different days. 
  Error bars: standard error. $\infty$: unconfined quasi-2D chambers.
  (b) Example of a correlation function $C(s)$ experimentally measured and fitted.}
\end{figure*}

We first compute the spatial correlation from which we extract two characteristic lengths: 
the oscillation period $L_c$ and the amplitude decay length $L_e$.
To do so, we take advantage of the lateral confinement and modify the classical two-point velocity correlation 
function to focus on the variation of the orthogonal flow $u_y$ along the channel length $\mathbf{e}_x$:
\begin{equation}
	C(s) = \frac{\sum u_y(x,y,t) \cdot u_y(x+s,y,t)}{\sum ||u_y(x,y,t)||^2},
\end{equation}
where $u_y = \mathbf{u}\cdot\mathbf{e}_y$ and taking the sum over PIV sub-windows in the bulk of a channel 
of a full $5$ second long movie. $C$ exhibits decaying oscillations (figure \ref{graph}(b)), a behaviour that was 
observed in previous work~\cite{Cisneros2007} but, to our knowledge, has not been fully analysed. 

Two main approaches have been used to compute the correlation length: either fitting with 
a decaying exponential~\cite{Sokolov2012, Gachelin2013} or 
taking the distance at which $C$ reaches zero or its first minimum~\cite{Wensink2008}. 
Instead we fit $C$ with a function that describes the decay and the oscillations:
\begin{equation}
	C_{\textrm{fit}} = \left[ A e^{-s/L_e} + (1-A) \right] \cdot \cos(2\pi s/L_c),
\end{equation}
where the first term indicates the amplitude decaying over the length $L_e$  
and the cosine term highlights the oscillation period $L_c$ (swirl size).

Figure \ref{graph}(a) shows that both $L_e$ and $L_c$ increase with the racetrack width and reach 
a plateau around $w\approx85\,\mu$m. This size is comparable with $w^*=70\,\mu$m when 
streaming ceases and the suspension is able to form full vortices. Moreover values at the plateau, 
$L^*_e\approx70\,\mu$m and $L^*_c\approx90\,\mu$m, are comparable in large racetracks and unconfined 
2D chambers (noted $\infty$ on the graphs).

Even though we also do observe a wave pattern in simulations, the correlation function mostly reflects the 
domain periodicity.

The bacterial flow patterns then are not static: while the bacteria stream in the channel, the partial 
swirls they generate travel as well. To quantify this effect we measure the flow pattern motion by the spatial 
correlation between two time points:
\begin{equation}
	C_{\mathrm{swirl}}(s) = \frac{\sum u_y(x,y,t) \cdot u_y(x+s,y,t+\Delta t)}{\sum ||u_y(x,y,t)||^2},
\end{equation}
taking the sum over all PIV sub-windows inside the channel and over a full $5$ second long movie. $C_{\mathrm{swirl}}$ takes a maximum value at $s^*$ which gives a wave speed: $U_{\mathrm{wave}}=s^*/\Delta t$.
We compare this speed to the averaged circulation speed $<u_x>_{x,y,t}$, and find that the wave travels slightly faster
than the bacteria: Median$\left(\frac{U_{\mathrm{wave}}}{<u_x>}\right)=1.2$ (standard error: $0.1$). Yet,
a more detailed analysis is required to fully understand the dynamics between cell motion and wave propagation in this particular confinement.

\begin{figure*}[htp]
  \centering
  \includegraphics[width = 0.95\textwidth]{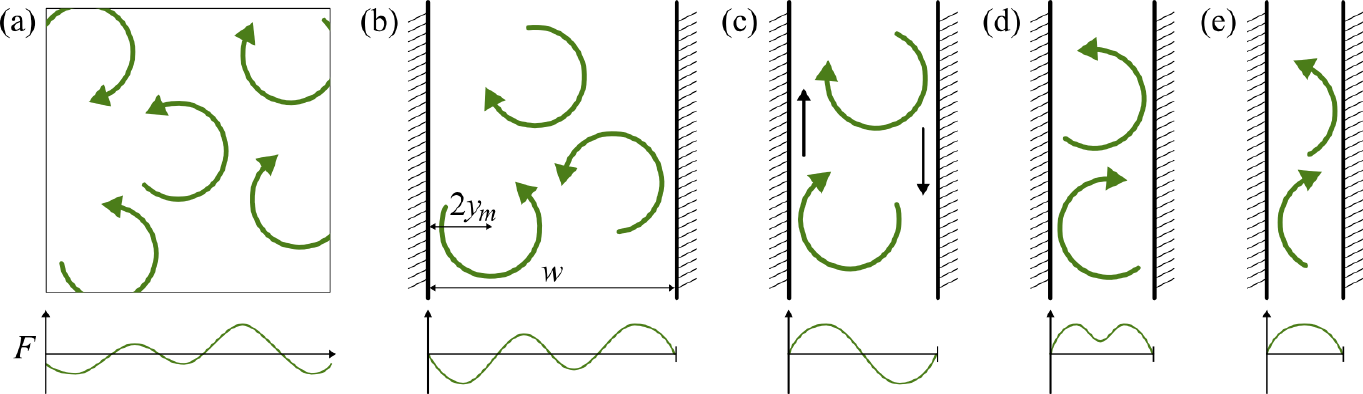}
  \caption{\label{sketch} Illustration of the bacterial organization in racetracks of varying width.}
\end{figure*}

%%%%%%%%%%
\section{Discussion}
We have shown that confining a dense bacterial suspension into a thin periodic racetrack leads to the 
spontaneous formation of a stable circulation along it.  
A similar setup has been previously used to study two different active matter systems: 
marching locusts~\cite{Buhl2006} and rolling colloidal particles~\cite{Bricard2013}. 
As for bacteria, collective colloid motion is driven by hydrodynamic interactions, 
but can result in the formation of travelling waves and density shocks while the bacterial 
suspension always appears spatially homogeneous. \\

To understand how the suspension self-organises in racetracks, we have measured the net flow, flow correlation 
and profile depending on the channel width. We reproduced these results with simulations, revealing the fluid flows
generated by the bacteria. Combining these different approaches, we draw a picture in Figure \ref{sketch} of how bacterial motion, 
hydrodynamic interactions and confinement drive the macroscopic circulation.

Independently of the channel size, bacteria at the PDMS surface move along a given direction while propelling
the fluid in the opposite direction, over a distance $y_m\approx14\,\mu$m. 
This fluid flow, much faster than the single cell's swimming speed, reorients and advects cells in the suspension. 
As a result bulk bacteria are transported against their swimming orientation in a backward motion.
Different behaviours then arise for varying channel widths.
For thin channels ($w<w^*=70\,\mu$m) the fluid flow on one side reaches the opposite interface and reorients 
the bacteria such that all cells at the different surfaces move in the opposite direction to the bulk circulation,
and give rise to the parabolic, flat and double-peaked profiles (figure \ref{sketch}).
Inside the channel, confinement restricts the turbulent bacterial motion such that bacteria form partial swirls 
that mostly travel with the bacterial stream, without affecting the overall circulation.
For intermediate channel widths ($70<w<90\,\mu$m) the suspension forms full vortices, showing no net circulation ($\Phi<0.2$).
Yet the coordination between opposite sides is not lost but now appears to be anti-correlated (as shown in the sinusoidal flow 
profiles, figure \ref{sketch}). In particular, vortices in the channels have a preferred rotation direction, either clockwise or 
counterclockwise, depending on the orientation of the bacteria at the interface.
For even larger channels, $w>90\,\mu$m, the suspension recovers its unconfined turbulent state 
and no correlation is observed between opposite PDMS walls.
However it still remains unclear how the critical length-scale of $w^*=70\,\mu$m of the vortical macroscopic structures in the collective motion of this bacterium is selected. \\

Some continuum models of active matter in confinement have previously 
predicted unidirectional flow or net circulation~\cite{Voiturez2005, Fielding2011, Woodhouse2012, Ravnik2013, Neef2014}.
Ravnik {\it et al.} \cite{Ravnik2013} found that the bacterial flow in a pipe has a weak 
($\sim1\%$) component along the $y$ and $z$ directions, but with vortex patterns quite different 
from the wave-like stream we observe here. Fielding {\it et al.}\cite{Fielding2011} considered 
2D channels of varying width but observed circulation for channels wider than the suspension vortex size, 
in contradiction with our net flow measurements (figure \ref{circulation}).  
Neef and Kruse~\cite{Neef2014} consider active polar fluids in annular domains and 
see surprisingly similar patterns to ours such as unidirectional streaming, 
moving or stationary vortices for increasing channel widths in the extensile particle case.
Many of the differences between these theoretical predictions
may arise from the difficulty of setting realistic boundary conditions
and on including both the swimmer-generated fluid flow and bacterial motion in such continuum models.

Despite many previous simulation studies on micro-swimmers in channels or between plates, 
this specific collective behaviour we report here has not been observed before in experiments or simulations.
Menzel~\cite{Menzel2013} performed simulations of self-propelled 
 particles using a Vicsek model with aligning interactions, no hydrodynamics, and observed collective 
migration along the channel in the form of particle clusters and lanes.
Costanzo {\it et al.}\cite{Constanzo2012} simulated elongated swimmers in a periodic 
channel, interacting directly and through dipolar-generated fluid flows, but did not 
observe any collective behaviour and net circulation when external flow was not imposed. 
One possible explanation is that they computed the hydrodynamic interactions 
without taking into account boundaries. Here we include mirror images to approximate the 
PDMS interface which enhances the bulk fluid stream. These small differences 
between these studies highlight the key roles played not only by the surfaces and the cell-driven flows 
arising from them but also of the elongated particle shapes that are needed for the 
swimmers orienting and ordering alongside each-other at the wall. 

Yet, the behaviour of bacteria in racetracks is reminiscent of some previous 
experiments and can be understood as a combination of  insights deriving from them. Swimming bacteria   
glued to a surface can coordinate their orientations and create a net flow far from the wall. 
Darnton {\it et al.}\cite{Darnton2004} used this effect to propel small objects while Kim 
{\it et al.}\cite{Kim2008} used it to turn a microfluidic channel into a bacteria-powered pump. This 
phenomenon is qualitatively similar to what we observed in the racetracks, except here the bacteria are free and
swimming along the surface while creating fluid flow in the bulk. 
Other studies have shown that the motion of swimmers in micro-channels is affected by an external 
shear flow~\cite{Rusconi2014, Hill2007, Kaya2012, Marcos2012, Kantsler2014}. In particular, swimmers there are biased to swim against 
the flow as we also observed in simulations and confirmed with fluorescence labelling. Finally, the self-organisation 
in racetracks is comparable to what was observed when bacteria were confined in flattened drops: the bacteria at the interface 
move in opposite direction to the bulk, which itself is advected by the fluid~\cite{Wioland2013, Lushi2014}. 
In both experiments (drops and racetracks), we found that the net circulation breaks down around $70\,\mu$m, 
the typical size of swirls in unconfined chambers~\cite{Dunkel2013} 
and also the critical diameter of circular drops below which bacterial motion stabilizes into one vortex~\cite{Wioland2013}.

Microfluidic channels have been used previously to study the behaviour of micro-swimmers, notably to sort bacteria {\it E. coli}
by length~\cite{Hulme2008} or direct spermatozoa~\cite{Denissenko2012, Guidobaldi2014} by 
making use of ratcheting channels. The curves of the micro-channels in those cases are designed to 
guide the swimmers in one specified direction. 
Here we used straight walls, such that the spontaneously emerging stream can occur in either direction with equal probability. 
As an extension of our work, one could design a circular channel with ratcheted surfaces to direct the suspension
stream in a chosen left or right direction. 

Dense bacterial suspensions have been extensively studied without confinement and yet their self-organisation
presents new challenges and surprises.
The experiments presented here not only give insight into the effect of confinement
but also help in understanding the general behaviour of such suspensions.
In particular, we have shown that the correlation function could be analysed as two parts: 
an oscillating and a decaying term. 
To our knowledge, other studies have described only one of the two, fitting the vortex size or 
the persistence length. Here confinement into a racetrack stabilises the suspension to form a more regular 
pattern, emphasising this structure. With the added insight of our work, some previous results could be reinterpreted: 
for example, Gachelin {\it et al.}\cite{Gachelin2013} have shown the persistence length of an \textit{E. coli} 
suspension to increase with the density, which translates to a slower decay of the correlation function $C(s)$ at higher densities.
However the first zero of the $C(s)$ occurred, over all their measurements, at the constant value $s^{\ast}\approx55\,\mu$m, 
indicating that the swirl size does not depend on the concentration 
but is an intrinsic length scale associated with the swimming bacteria.

Finally, this confinement setup opens new avenues for the study of collective behaviours in more complex natural
or artificial environments, that could include networks of channels, various solid or fluid interfaces, and 
chemical attractants. Notably, Wilking {\it et al.} showed \textit{B. subtilis} biofilms to form channels, 
on average $90\,\mu$m in diameter~\cite{Wilking2013}, comparable with our critical width $w^\ast\approx 70\,\mu$m. 
Although the primary role of these channels is to transport liquid and nutrients, they could also 
direct the migration of single or groups of swimming bacteria.
A better understanding of the interactions and how they drive phenomena in these systems 
can guide us into better control of collective bacterial behaviour and possible use in technological applications.

\section*{Acknowledgments}
%\ack
This work was supported in part by the European Research Council Advanced Investigator Grant
247333 (H.W. and R.E.G.) and the National Science Foundation grant CBET-1544196 (E.L).

%\endbib

\end{document}